\documentclass[10pt, conference, compsocconf]{sty/IEEEtran}
\ifCLASSINFOpdf
\else
\fi
\usepackage{graphicx}
\usepackage[ruled,vlined,linesnumbered]{algorithm2e}
\usepackage{multirow}
\usepackage{listings}
\usepackage{booktabs}
\usepackage{color}
\usepackage{url}

\begin{document}
%
\title{Dynamically Provisioning Cray DataWarp Storage}


\author{\IEEEauthorblockN{Fran\c{c}ois Tessier, Maxime Martinasso, Matteo Chesi, 
Mark Klein, Miguel Gila}
\IEEEauthorblockA{Swiss National Supercomputing Centre, ETH Zurich, Lugano, Switzerland\\
  \{firstname\}.\{lastname\}@cscs.ch}
}


%


\maketitle

\begin{abstract}

Complex applications and workflows needs are often exclusively expressed in terms of computational resources on HPC systems. In many cases, other resources like storage or network are not allocatable and are shared across the entire HPC system. 
By looking at the storage resource in particular, any workflow or application should be able to select both its preferred data manager and its required storage capability or capacity. To achieve such a goal, new mechanisms should be introduced.
In this work, we introduce such a mechanism for dynamically provision a data management system on top of storage devices. We particularly focus our effort on deploying a BeeGFS instance across multiple DataWarp nodes on a Cray XC50 system. However, we also demonstrate that the same mechanism can be used to deploy BeeGFS on non-Cray system.
\end{abstract}

\begin{IEEEkeywords} Dynamic provisioning; Storage; Data manager; DataWarp; BeeGFS;

\end{IEEEkeywords}

%
\IEEEpeerreviewmaketitle

\section{Introduction}
\label{sec:introduction}

Large and complex scientific workflows such as generation of weather
forecast data~\cite{kronos} or identification of new materials~\cite{PIZZI2016218} define a set of tasks
and dependencies among themselves. When running on a large and shared
HPC system, these workflows are expressed in terms of discrete compute
tasks whereas data management relies on accessing a global
shared file system. Workflow data capability is therefore constrained by the
parallel file system both for data format and performance
variability. From a scientist point of view, expressing data-oriented
tasks inside workflows enables a
greater flexibility and a more complete definition of the workflow
itself. However, on the HPC center side, it is not feasible to create
an HPC system supporting a large variety of data manager systems such
as parallel file systems or databases~\cite{JSFI162}. Even more due to the economy of
scale of resources, HPC centers tend to provide shared data resources
(but exclusive compute resources).

The past years have seen the amount of data generated by
scientific workflows and large-scale HPC simulations dramatically
increase. Despite attempts by vendors to temper this burden by
deploying new tiers of memory and storage, it is clear that the
performance gap between computing power and I/O operations continues to grow. 
Burst buffers for instance, such as Cray DataWarp~\cite{HenselerDW}, or
hybrid storage tiers, such as NVMe, have been designed to mitigate the
I/O slowdown by providing an intermediate tier of fast storage
between the compute nodes and the parallel file system.
In the context of HPC storage, the
multiplication of layers in the I/O software stack (specialized stack
for shared resources, databases over file system, etc...) limits the capability to access the full potential of the I/O hardware performance.
Application developers are
dependent on the software stack deployed on the system.
For example,
the Cray Data Virtualization Service (DVS) is necessary to use
DataWarp nodes.
On-node disk is another example of storage layer whose
use is limited to the deployed file
system. 

In this work, we propose to dynamically provision HPC storage resources for workflows and applications.
As for computing resources, data resources are managed as a batch scheduling resource and are requested as the job submission of the application or workflow task. The selection of the deployed data manager is done inside the job scripts.
The flexibility offered by such a dynamic provisioning mechanism is the key here.

As a concrete example, we focus our study on dynamically deploying BeeGFS~\cite{beegfs} on a set of Cray DataWarp nodes.
We first repurpose DataWarp nodes with compute node images
which allows us on one hand to configure DataWarp as an allocatable
resource in Slurm and on the other hand to configure the raw storage
devices for any data managers. In a second step, we enable the
capability to deploy on-demand a well-sized BeeGFS on those DataWarp
storage devices. Moreover, we show the portability of our method on non-Cray solutions hosting NVMe disks. 
We validate the reliability of this architecture by subjecting
it to a high I/O load through benchmarks representative of the typical
workloads of applications running on HPC systems.

The key contributions of our work are:
\begin{itemize}
    \item to introduce a simple and portable mechanism to deploy on-demand data managers;
    \item to propose a new usage of Cray DataWarp as an I/O storage for on-demand deployment of the BeeGFS file system;
    \item to validate the performance of the BeeGFS file system deployed in this way.
\end{itemize}
\section{Context and Motivation}
\label{sec:context}

Traditional HPC centers focus on providing one highly performing flagship machine for a precise set of scientists of different domains. Oppositely, Cloud providers intend to give to anyone access to commodity computation and storage capability. Because of their extremely general user base, they have developed infrastructure-as-a-service (IaaS) technology to let users configure and deploy the system they require. One key element of the IaaS technology is the dynamic provisioning of resources: compute, network and storage. 

It is common for HPC system to provide dynamic access to compute nodes through a batch scheduler, however, little has been done for dynamically provisioning storage and network resources. Such resources are traditionally shared among all users. To maximize performance many HPC techniques have been developed to minimize contention and congestion on these two resources. By taking an IaaS approach, in this work, we introduce the idea of dynamically provisioning storage resources on contemporary HPC hardware.

There are many advantages for dynamically provisioning storage resources in an HPC context. For workflows and applications it allows to define precisely data managers type and configuration. It also allows to select the storage hardware to use. It brings isolation by limiting access to the data managers to the application or workflow. Storage resources are not shared anymore among users. For instance, a metadata sensitive application could deploy a fast metadata file system on close-to-compute flash storage~\cite{kim2008efficient} to maximize performance and to avoid metadata contention initiated by other applications running on the system.



%


\section{Architecture and Implementation}
\label{sec:architecture}

Our dynamic resource provisioning mechanism consists of deploying on demand, a data management system (currently a parallel file system, but also supporting object-based storage or databases in the future) on raw intermediate storage. To do so, we identified a list of requirements for accessing the underlying storage and we developed our provisioning mechanism such as resources can be seamlessly requested through the job scheduler. 

In this section, we will first present the main architecture of our dynamic resource provisioning mechanism. Then, we will list the prerequisites needed to use the intermediate storage layer. Finally, We will give implementation details regarding the deployment of BeeGFS on top of Cray DataWarp nodes. For the rest of this paper, we will use the term \emph{storage node} to talk about nodes with local storage, as opposed to compute nodes. The set of storage nodes and the set of compute nodes can be disjoints (Cray DataWarp nodes) or the former can be a subset of the latter (node-local storage).

\subsection{Dynamic resource provisioning}

\begin{figure}[h]
  \centering
  \includegraphics[width=0.98\linewidth]{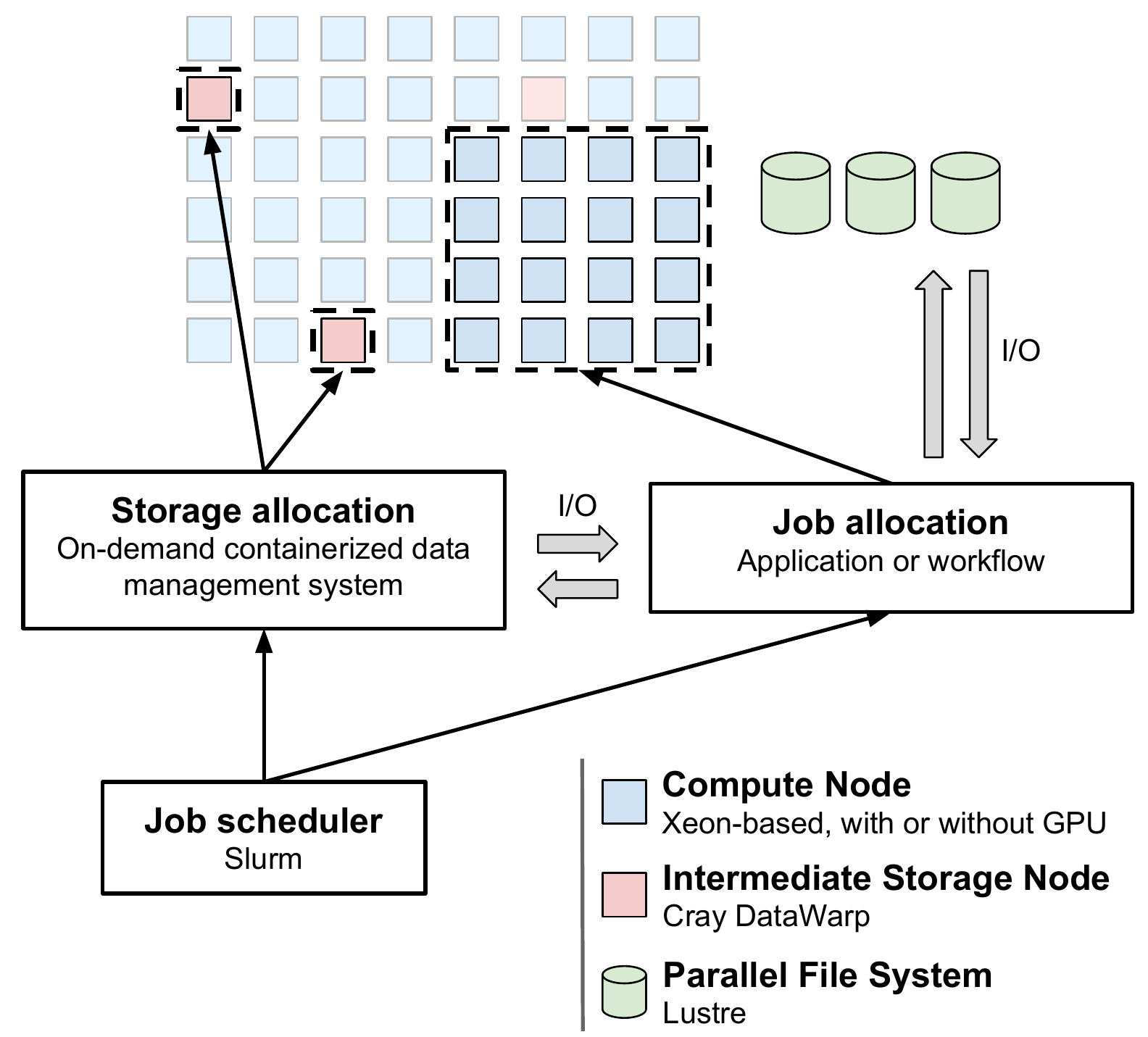}
  \caption{On-demand data management system}
  \label{fig:dyn_pro}
\end{figure}

Figure~\ref{fig:dyn_pro} depicts how our dynamic resource provisioning mechanism works on a HPC system. Through the job scheduler, a user can request two allocations: the compute nodes needed to run the application(s) and a set of storage nodes to deploy a data manager. We use SLURM~\cite{10.1007/10968987_3} as a job scheduler here. However, it is important to note that our approach is independent of the job scheduler as long as it is possible to request an allocation of storage nodes. Once allocations have been granted, data management services are deployed to the storage nodes and clients are set up on the compute nodes.

On the storage allocation side, containers packaging the data management system are started on the nodes. A script is launched along with the container then configures and starts the services. A containerized approach generally allows the required software stack to start the services without superuser privileges. Those services will remain accessible from outside of the container as the processes are visible in the PID namespace of the host. 

On the compute node side, it is necessary to configure access to the previously deployed data manager. Depending on the data management system, this step can be done by simply giving the master node's IP address or by a more complex manner with a daemon, a kernel module or a container if necessary.

As shown on Figure~\ref{fig:dyn_pro}, this architecture offers two options to the running application to perform I/O. While the shared parallel file system is still accessible, a temporary data manager can now be accessed.

When the computing and storage resources are released by the user or the job scheduler, services on storage nodes are killed and data on disks is deleted. On compute nodes, clients are properly stopped.

\subsection{Prerequisites for accessing intermediate storage}
\def\code#1{\texttt{#1}}
Depending on the type of intermediate storage targeted, system administration may be required to allow the deployment of an on-demand data manager. We make here the assumption that nodes (storage and compute) are part of the same network and can be mutually reachable.

On a typical HPC system with node-local storage, disks are usually directly writable. In the case of DataWarp, however, intermediate storage is distributed across dedicated nodes and accessed through the DVS layer. DVS (for Data Virtualization Service) does the projection of DataWarp storage onto the compute nodes.
In the standard implementation of DataWarp end users cannot access DataWarp nodes but, instead, can only interact with their projected storage resources.
In order to grant a higher level of interaction with DataWarp nodes and their resources, we have re-purposed them from hidden service nodes to standard compute nodes with just a minimal system customization layer to setup their local NVMe storage. 
The re-purposing consisted in two simple reconfiguration changes in the Cray XC system configuration database. We first modified the node type from \code{service} to \code{compute} through the \code{xtprocadmin} command then we mapped a compute node image to boot with thanks to the \code{cnode update} CLI tool.
The additional configuration changes needed were made through Ansible, a popular system administration tool for distributed deployment, and consisted in formatting flash devices with a XFS file system and mount them on the node with all permissions granted to any user.

While storage embedded within compute nodes is accessible with a standard allocation, DataWarp nodes allocation must be requested and specified by the user using SLURM Burst Buffer plugin interface.
This interface is well suited for a standard Burst Buffer implementation to do check-pointing, but it limits the user interaction with storage resources to those functionalities already coded in the interface. A strong coding effort is required to change or enhance this plugin.

In order to overcome these limitations imposed by SLURM Burst buffer interface, the DataWarp nodes, already re-purposed as compute nodes with local flash storage, have been made available to end users through a SLURM constraint. In the same way users can use \code{gpu} constraint to request nodes equipped with GPU or \code{mc} constraint for a multicore compute node  on CSCS systems, the storage nodes are provided requesting \code{storage} constraint.

\subsection{Implementation details}
The proof of concept of our dynamic provisioning method has been implemented as bash scripts deploying python scripts and containerized software stack. For each allocation (storage and compute), a bash script performs a loop on the list of granted nodes. On storage nodes, a Docker container~\cite{Merkel:2014:DLL:2600239.2600241} is started with Shifter~\cite{DBLP:journals/corr/BenedicicCMM17,gerhardt:shifter}. The list of available disks and their mount points of each node are described in a configuration file which is included inside the container. In the future, we plan to dynamically modify this configuration file.
On the compute nodes, a python script is executed on each node. Those two steps can be carried out with a batch script or within an interactive session on nodes. 

To validate our dynamic provisioning method, we used BeeGFS as a data manager. BeeGFS is a POSIX-compliant parallel file system with a client-server architecture. It features five main components:
\begin{itemize}
    \item A management server in charge of orchestrating the other daemons;
    \item at least one metadata server for metadata;
    \item at least one storage server for raw data;
    \item A monitoring service accessible from a desktop Java application;
    \item A BeeGFS kernel module for the client.
\end{itemize}

For our study using BeeGFS, the container deployed on storage nodes contains a fresh install of all the packages required to start the parallel file system. A Python script is set as the entry point of the container, meaning that it is started at container launchtime. This script is in charge of creating all the configuration files for each server-side component of BeeGFS: management, metadata, storage and monitoring. Particularly, it sets up the network parameters (IP of the management server, communication ports), the absolute path of the mount point of the disk that will be used by the service (\emph{/mnt/nvme0n1} for metadata for instance) and a few other daemon-specific settings like, for example, the capability to use file-system extended attributes for metadata. This script finally starts all the daemons within the container in user-space. 

On the compute nodes, a script initializes the BeeGFS client configuration and creates a local mount point of the running BeeGFS instance. This last step has a limitation we will need to address in the future: a kernel module is necessary to mount the file system implying system administration privileges to be installed. In addition, special privileges have to be granted to the user (hence, the script) on the operating system to be authorized to locally mount BeeGFS. On the experimentation platforms we used to evaluate our dynamic resource provisioning mechanism, privileges have been escalated to let us configure this setup. To overcome these limitations on production systems, we plan to investigate a solution where is the job scheduler during prolog execution to setup the environment as the user requested, loading the kernel module and then building and mounting the file system. In the same way, we plan that at the end of the job, during epilog execution, the job scheduler will unmount and delete the file system and then unload the kernel module.

Once those constraints have been mitigated our client-side script can execute the following command on the compute nodes:

\lstset{basicstyle=\footnotesize}
\begin{lstlisting}[language=bash]
$ mount -t beegfs beegfs_nodev \ 
        <mount_point> \
        -ocfgFile=beegfs-client.conf,_netdev,,
\end{lstlisting}

From an application point of view, writing or reading data to/from the local BeeGFS mount point is the way to query the parallel file system.

\section{Performance}
\label{sec:performance}

We present in this section a performance evaluation of the proposed solution on two systems. First, we targeted a Cray system with DataWarp nodes. We ran on this machine multiple tests with IOR~\cite{ior} to cover typical I/O workloads. We also carried out experiments with HACC-IO~\cite{HABIB201649}, the I/O kernel of a large-scale cosmological application. Secondly, in order to show the portability of our approach, we ran experiments on a compute node equipped with local NVMe disks.


\subsection{Cray XC50 with Cray DataWarp}

We first deployed our dynamic provisioning mechanism on Dom, a Cray XC50 system. Dom is the test and development system of Piz Daint, a 27 PFlops XC50 supercomputer at CSCS. The testbed features 8 nodes, each with two 18-core Intel Broadwell CPUs (Intel Xeon E5-2695 v4) and 64 GB of DRAM. Within the Cray Aries network connecting the compute nodes, we also find 4 DataWarp nodes each embedding three 5.9TB PCIe SSD (Samsung PM1725a) whose vendor's value for I/O bandwidth is 6.3GBps for sequential read and 2.6GBps for sequential write. Our experiments using the \code{dd} tool with multiple concurrent streams showed empirical peak performance values for reading and writing of respectively 6.34 and 3.2GBps. The global storage system on Dom provides 170TB of usable space managed by a Lustre file-system~\cite{lustre} and distributed across 2 OSTs (object storage target). Given the small scale of this cluster, we can consider Lustre as a dedicated parallel file-system.

For all of our experiments on Dom, we used two disks per DataWarp node for storage and one disk for metadata. The disk dedicated to metadata on the first node of the storage allocation was also used for BeeGFS management and monitoring. For Lustre, we set the stripe count (number of OST used to stripe files across) to 2. For the benchmarks presented below, we used the 8 compute nodes with 36 processes per node (288 processes total) for all the runs. The stripe size for both file-systems was set to 1MB.

\subsubsection{Deployment time}
The time needed to deploy a containerized and dynamically provisioned file-system on multiple nodes is difficult to really estimate. Lots of factors can intervene like the network connection between the management server and the metadata and storage daemons as well as the container runtime system. Our experiments, however, showed an average deployment time of 5.37 seconds over three runs for a deployment of BeeGFS on two DataWarp nodes.

\subsubsection{IOR}
\label{sssec:ior_dom}
The IOR~\cite{ior} suite is a popular and highly tunable set of I/O benchmarks, especially used for the IO-500 ranking~\cite{io-500:Release1.1}. It is composed of \emph{IOR} for evaluating I/O performance and \emph{mdtest} for appraising metadata management. We evaluated these two metrics on the on-demand BeeGFS file-system and on Lustre. For IOR, we focused our experiments on two ways of writing data out: a single shared file or one file per process. As our goal was to show how the file-systems can mitigate the burden caused by multiple streams, we performed independent MPI-IO calls instead of collective operations. We also set flags to disable client-side cache effect. We ran all the experiments ten times.

\begin{figure}[h]
  \centering
  \includegraphics[width=0.98\linewidth]{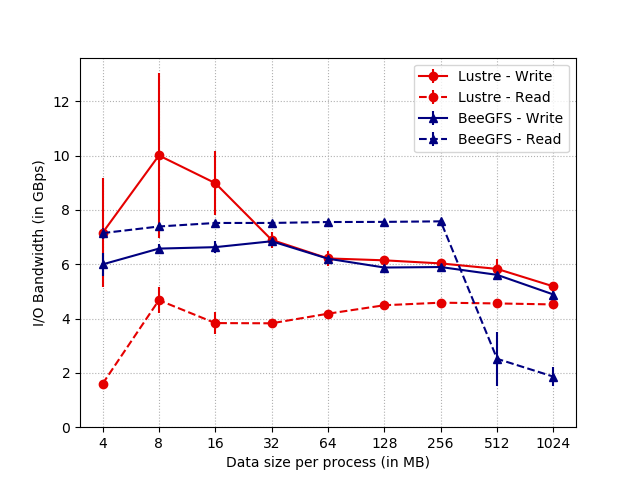}
  \caption{I/O bandwidth achieved on Dom with the IOR benchmark running on 8 compute nodes (36 ppn) according to the data size written per process into a \textbf{single shared file}. Comparison between on-demand BeeGFS deployed across 2 Cray DataWarp nodes and Lustre with 2 OSTs.}
  \label{fig:ior_dom_ssf}
\end{figure}

Figure~\ref{fig:ior_dom_ssf} compares the I/O bandwidth attained when the 288 processes write then read back a single shared file to/from the on-demand BeeGFS deployed over two DataWarp nodes and the Lustre file-system. We first observe than the write bandwidth is comparable from 32MB per process written into the shared file and beyond. Both file-systems achieve around 6GBps. With smaller sizes however, Lustre outperforms BeeGFS but at the cost of a significant variability. When reading back data, BeeGFS on two DataWarp nodes performs approximately 2x better than Lustre and even more with 4MB per process. Nevertheless, when reading back 512MB or 1GB per process from our on-demand BeeGFS, the read bandwidth dramatically decreases. We explain this behavior by the fact that BeeGFS caching mechanism size is limited to the 64GB of DRAM on each DataWarp node. With a balanced I/O load on the two nodes used here to deploy BeeGFS, the amount of data to manage per storage node is $\frac{1}{2} \times \#computenodes \times \#ppn \times S_p$, $S_p$ being the size of the data written or read per process. For $S_p\ge$512MB, this value is greater or equal to 73.72GB and the cache cannot fit in the available memory. The benefit of a server-side caching is therefore highly reduced.

\begin{figure}[h]
  \centering
  \includegraphics[width=0.98\linewidth]{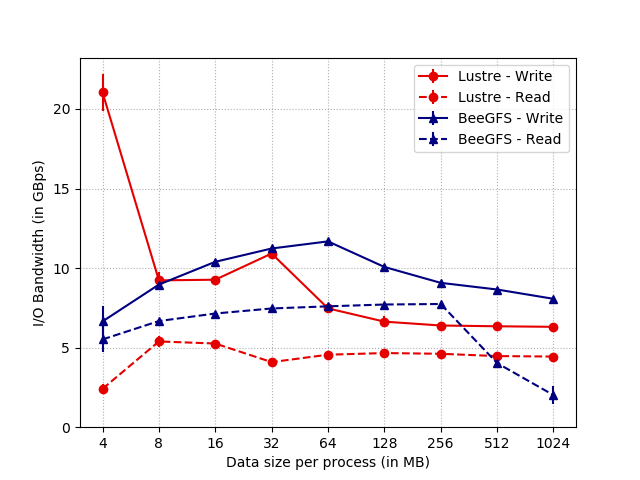}
  \caption{I/O bandwidth achieved on Dom with the IOR benchmark running on 8 compute nodes (36 ppn) according to the data size written per process into its own file (\textbf{one file per process)}. Comparison between on-demand BeeGFS deployed across 2 Cray DataWarp nodes and Lustre with 2 OSTs.}
  \label{fig:ior_dom_fpp}
\end{figure}

We depict on Figure~\ref{fig:ior_dom_fpp} the same experiments with one file per process written and read back instead of a single shared file. This file distribution is known to provide better I/O performance. Except for a few data sizes, the dynamically provisioned BeeGFS reaches an higher I/O bandwidth than Lustre for both writing and reading. When reading back data, we observe a similar behavior as described previously: with large sizes (512MB and 1GB per process), the I/O bandwidth is low. An unexpected behavior appears on Lustre when writing 4MB files. This level of performance cannot be achieved in practice, which leads us to believe that a write cache effect that we have not been able to contain is at work for this particular data size.

Another remark is that the peak write bandwidth recorded with BeeGFS (with 64MB/process) is 70\% higher than the peak bandwidth observed on a single shared file (11.96GBps against 7.01GBps). If we consider the sum of the write performance we observed of the four storage disks used for those experiments ($4 \times 3.2$GBps$ = 12.8$GBps), and given the management of metadata by two daemons on dedicated disks, we can conclude here that the file system is being used at the maximum of its capability.

\begin{figure}[h]
  \centering
  \includegraphics[width=0.98\linewidth]{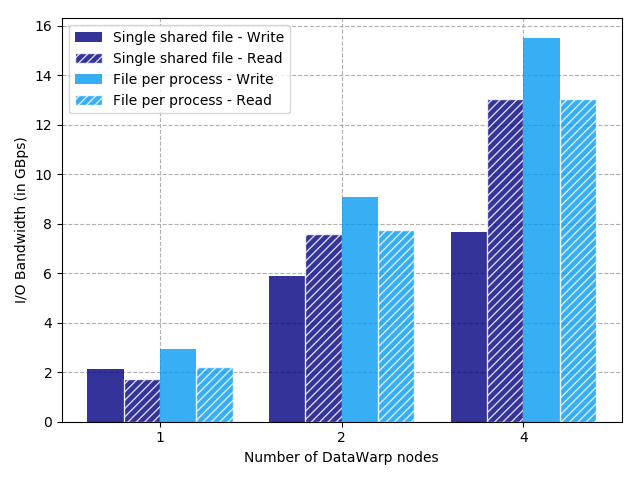}
  \caption{I/O bandwidth achieved on Dom with the IOR benchmark running on 8 compute nodes (36 ppn) while varying the size of the dynamically deployed data manager.}
  \label{fig:ior_dom_dw}
\end{figure}

The scalability of a dynamic resource provisioning system depends on multiple factors, both in terms of underlying hardware (number of nodes and disks) and software layer. 
Although our resources were limited we tried to evaluate this criteria. To do so, we ran the IOR benchmark from the 8 compute nodes while varying the size of the on-demand BeeGFS from 1 node to 4 nodes. We kept the ratio metadata:storage servers per node to 1:2. Figure~\ref{fig:ior_dom_dw} presents those results for both types of file distribution. As expected, the scalability is satisfying for almost all the cases. We could mention the write bandwidth with a single shared file whose performance improvement follows a logarithmic curve: the write bandwidth almost triples from 1 to 2 DataWarp nodes but is increased by only 30\% when doubling again the number of storage nodes.

\subsubsection{mdtest}
\label{sssec:mdtest_dom}
The \emph{mdtest} benchmark has been designed to evaluate the metadata performance of a parallel file-system. Directories and files are created, stated and removed multiple times and the number of operations per second is measured for each operation. Table~\ref{tab:mdtest_dom} shows the results of this benchmark on the BeeGFS over two DataWarp nodes and on the Lustre file-system. Except when stating a directory, Lustre metadata management outperforms the dynamically provisioned BeeGFS. File creation for instance is 3.5x faster on Lustre. The value obtained with BeeGFS for directory stat looks very high. A client-side cache probably explains this result. 

\begin{table}[h]
\centering
\begin{tabular}{@{}clrr@{}}
\cmidrule(l){3-4}
\multicolumn{1}{l}{}       &                                        & \multicolumn{1}{c}{\textbf{BeeGFS}} & \multicolumn{1}{c}{\textbf{Lustre}} \\ \midrule
\textbf{Target}            & \multicolumn{1}{c}{\textbf{Operation}} & \multicolumn{2}{c}{\textbf{Ops}}                                          \\ \midrule
\multirow{3}{*}{Directory} & creation                               & 8276.43                             & 37222.57                            \\
                           & stat                                   & 5301788.76                          & 182330.42                           \\
                           & removal                                & 12967.02                            & 38732.00                            \\ \midrule
\multirow{4}{*}{File}      & creation                               & 6618.37                             & 22916.15                            \\
                           & stat                                   & 144410.46                           & 169140.32                           \\
                           & read                                   & 22541.08                            & 45181.55                            \\
                           & removal                                & 8431.71                             & 35985.96                            \\ \midrule
\multirow{2}{*}{Tree}      & creation                               & 2183.40                             & 3310.42                             \\
                           & removal                                & 125.23                              & 1298.55                             \\ \bottomrule
\end{tabular}
\caption{I/O operations per second for various operations performed on the dynamically provisioned BeeGFS and on Lustre with the mdtest benchmark on Dom from 8 nodes (36 ppn).}
\label{tab:mdtest_dom}
\end{table}

\subsubsection{HACC-IO}

\begin{figure}[h]
  \centering
  \includegraphics[width=0.98\linewidth]{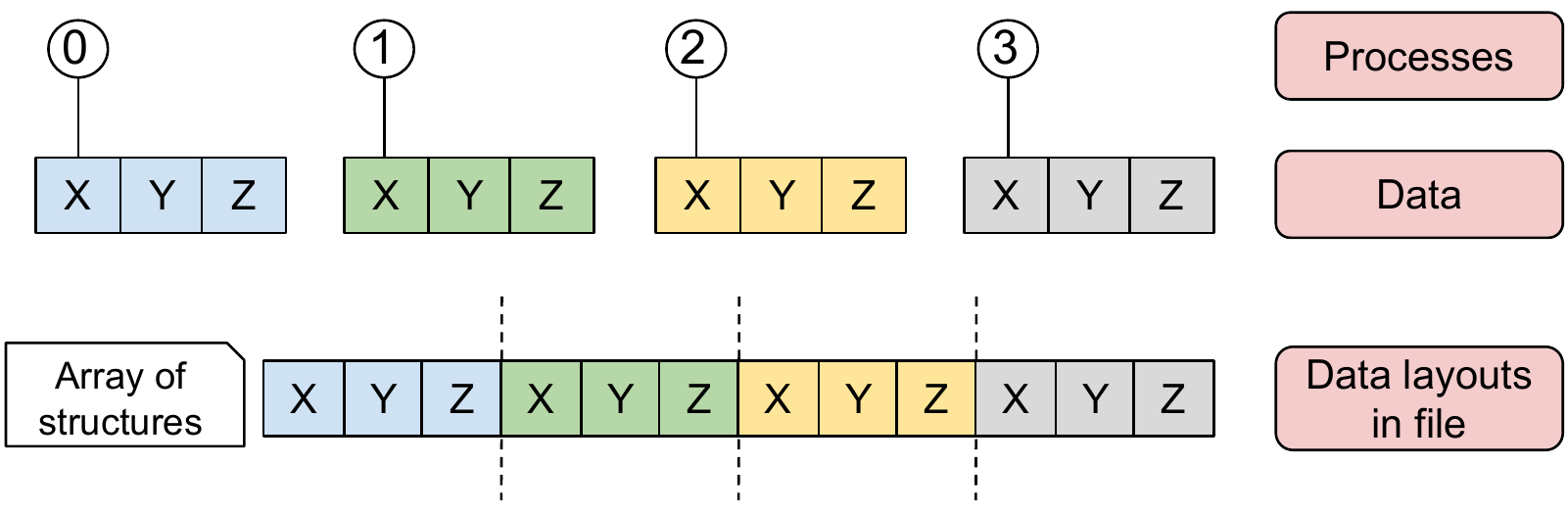}
  \caption{Array of structure data layout implemented in HACC-IO}
  \label{fig:hacc_aos}
\end{figure}

HACC-IO is the I/O kernel of HACC~\cite{HABIB201649} (Hardware Accelerated Cosmology Code). This large-scale cosmological application, developed at Argonne National Laboratory, requires the massive compute power of supercomputers to simulate the mass evolution of the universe with particle-mesh techniques. In terms of I/O, every process of a HACC simulation manages a number of particles. Each particle is defined by nine variables---$XX$, $YY$, $ZZ$, $VX$, $VY$, $VZ$, $phi$, $pid$, and $mask$---corresponding to the coordinates, the velocity vector, and relevant physics properties. The size of a particle is 38 bytes. A useful base value of 25,000 particles requires approximately 1 MB. The data layout of those particles in file follows an array of structure pattern as described in Figure~\ref{fig:hacc_aos}. Data is written and read back from a single shared file by all the processes.

\begin{figure}[h]
  \centering
  \includegraphics[width=0.98\linewidth]{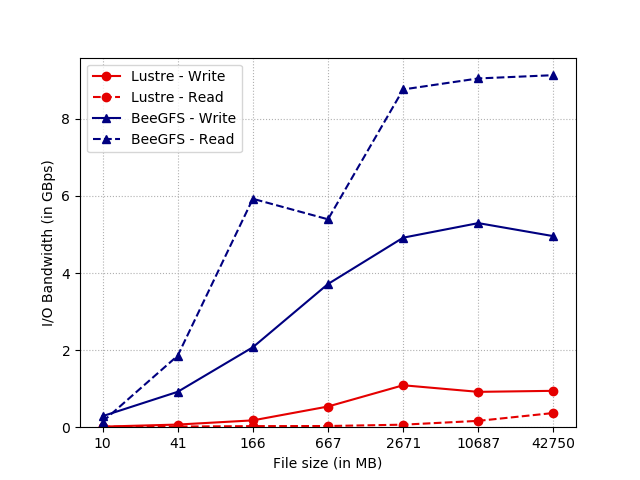}
  \caption{I/O bandwidth achieved on Dom with the HACC-IO running on 8 compute nodes (36 ppn) according to the number of particles (data size) per process into a \textbf{single shared file}. Comparison between on-demand BeeGFS deployed across 2 Cray DataWarp nodes and Lustre with 2 OSTs.}
  \label{fig:hacc_dom}
\end{figure}

Similarly to the previous experiments, we ran HACC-IO on 8 compute nodes (36 ppn) on Dom and compared the I/O performance of the two file-systems, BeeGFS and Lustre, respectively using two DataWarp nodes and two OSTs. We see on Figure~\ref{fig:hacc_dom} that the on-demand file-system offers the best read and write bandwidth up to a 42GB file size. The peak write bandwidth is 5.3GBps while data is read back up to 9.1GBps. The Lustre file-system, however, performs poorly. 1GBps is barely attained during the write phase. The read bandwidth stays below 0.4GBps regardless the input file size. Previous work~\cite{Tessier:2018:ODA:3205289.3205316} evaluating HACC-IO on Lustre corroborates those results.

\subsection{Non-Cray testbed}

To assess the portability of the proposed approach, we conducted experiments on another system, called Ault. Ault is testbed platform at CSCS that allows for prototyping experimental services and platforms. Various types of hardware is available for researchers to quickly provision as needed using Canonical's Metal as a Service (MaaS) product. This allows for safe privileged-access level experimentation by researchers without impacting production services. For our experiments, we used Ault11, a compute node with a 22-core Intel Xeon Gold 6152 CPU cadenced at 2.10GHz. The node also hosted 16 NVMe disks (Intel SSD DC P4500) whose vendor's value for sequential read and write is respectively 3.2GBps and 1.9GBps. As for the disks on DataWarp nodes, this performance values do not reflect a real use-case with multiple concurrent streams.

We set up the following configuration for the on-demand file-system:
\begin{itemize} 
  \item 1 disk for BeeGFS management and monitoring
  \item 2 disks for the metadata
  \item 5 disks for distributed storage
\end{itemize}


\subsubsection{Deployment time}
On Ault, the deployment time for the BeeGFS instance across 8 empty disks is approximately 4.6 seconds. If the tree structure of the file system has been written already, this initialization phase changes to 1.2 seconds.

\subsubsection{IOR}
\begin{figure}[h]
  \centering
  \includegraphics[width=0.98\linewidth]{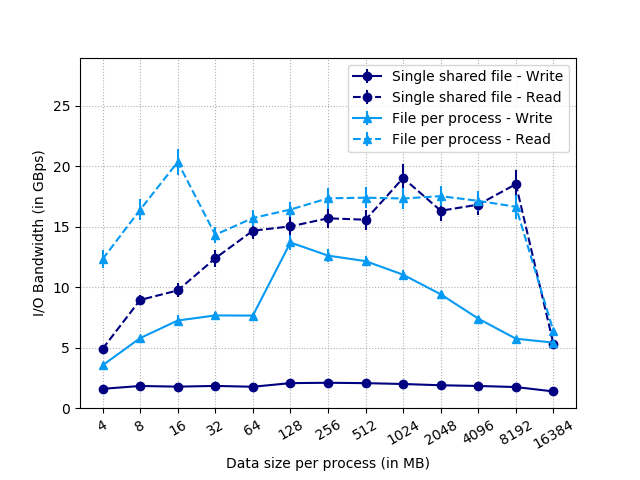}
  \caption{I/O bandwidth achieved on Ault with the IOR benchmark running on one compute node (22 ppn) according to the data size written per process. The on-demand BeeGFS is deployed on 8 NVMe disks.}
  \label{fig:ior_ault}
\end{figure}

We ran on this platform the same IOR experiments as in~\ref{sssec:ior_dom} adapted to the number of available computing and storage resources. Figure~\ref{fig:ior_ault} depicts the results of the IOR benchmark on the node-local on-demand BeeGFS with a single shared file and one file per process. Results are in correlation with the empirical performance we can expect for those disks with multiple concurrent streams. The peak read bandwidth attained 20.36GBps following a file-per-process distribution while, for the same file division, the peak write bandwidth reached 13.70GBps. Again, performing I/O into dedicated files per process substantially increases the I/O bandwidth.

\subsubsection{mdtest}
Table~\ref{tab:mdtest_ault} presents the metadata performance of the on-demand BeeGFS distributed across 8 NVMe disks on Ault. 

\begin{table}[h]
\centering
\begin{tabular}{@{}clr@{}}
\toprule
\textbf{Target}            & \textbf{Operation} & \textbf{Ops} \\ \midrule
\multirow{3}{*}{Directory} & creation           & 1796.31                 \\
                           & stat               & 667250.43               \\
                           & removal            & 5516.92                 \\ \midrule
\multirow{4}{*}{File}      & creation           & 5234.87                 \\
                           & stat               & 98888.28                \\
                           & read               & 22889.51                \\
                           & removal            & 5929.99                 \\ \midrule
\multirow{2}{*}{Tree}      & creation           & 2754.81                 \\
                           & removal            & 980.84                  \\ \bottomrule
\end{tabular}
\caption{I/O operations per second for various operations performed on the dynamically provisioned BeeGFS with the mdtest benchmark on Ault from one node (22 ppn).}
\label{tab:mdtest_ault}
\end{table}
\section{limitations}
\label{sec:limitations}

Our proposed approach for doing dynamic provisioning suffers from a set of limitations, in particular, by using a kernel-space file system such as BeeGFS.

\begin{itemize}
    \item When a storage is dynamically provisioned by a job, it does not contain any data. Therefore, a stage in and stage out of data might be required for the scientific application to run or to retrieve its results. Nevertheless, many high-end HPC systems offer a high-performance scratch file system for which such stage in and out steps are necessary. The only difference is that the movements of data in the latter case are not accounted in the job runtime.
    \item As mentioned, a kernel-space file system like BeeGFS requires privileges to be able to run (kernel module, mount command). Whereas, creating a file system and mounting it could be done by prolog script and executed by Slurm (avoiding the need to escalate privileges at the user level), the node images need to have the kernel modules pre-installed. 
    \item The notion of resources for storage can be either focusing on capacity (quantity of bytes) or capability (speed of read/write operations). In the later case, using more disks increases the overall I/O operation bandwidth. It means that if storage capability is targeted by the user, he should ask for more storage nodes possibly wasting disk capacity. In any case, many compute jobs also under-utilize their compute resources by using only a portion of the node memory or efficiently utilizing either an accelerator or the host processor.
\end{itemize}
\section{Related Works}
\label{sec:related}
The Cloud community has introduced the concept of IaaS. Technology such as 
OpenStack~\cite{Shrivastwa:2016:OBC:3099715} or 
Kubernetes~\cite{Hightower:2017:KUR:3175917} allows to create virtual clusters on hardware resources.
Even if both tools are technically different, their goals is to define virtual clusters by grouping amount of compute, network and storage resources. Storage is one of the resources, for OpenStack it could be either a block storage, an object storage, or ephemeral storage. Kubernetes increased the range of storage with a large list of possible options. However, none of them are integrated HPC storage file systems such as Lustre~\cite{Schwan03lustre:building} or IBM Spectrum Scale~\cite{schmuck2002gpfs} (formerly GPFS).

In parallel, Cloud providers are starting to propose HPC storage capability. Amazon Web Services (AWS)~\cite{Sammons:2016:IAB:3126334} is  leading this effort. Recently, AWS allows to deploy on-demand the Lustre file-system~\cite{lustreaws}. AWS also provides Spectrum scale~\cite{gpfsaws}, however, such deployment requires human intervention. By referring to AWS features, traditional HPC file systems are not solely limited to a one-time global installation per systems but have the possibility to be dynamically provisioning.

Some HPC-oriented file systems provide on-demand provisioning feature. BeeGFS~\cite{beegfs} provides the BeeOND~\cite{herold2014introduction} option. By using a single script a full instance of the file system can be instantiated. It appears that such script is very similar to our work to deploy BeeGFS, it follows the same steps but provides less flexibility and generality.

It is interesting to note that the idea of temporary file system on burst buffer~\cite{8514892}~\cite{7877147} have already been developed. However, such work focus on providing a new file system build with the capacity to be deployed on-demand on a burst-buffer device. Our work provides a more generic solution in term of data managers and target storage hardware.

\section{Conclusion}
\label{sec:conclusion}
We presented in this paper a proof of concept of a mechanism to dynamically provision a data manager on top of intermediate storage resources. Such an approach allows an application or workflow to get an isolated data management system meeting its requirements. As an example, we focused on deploying BeeGFS, a parallel file system, over intermediate storage.

We evaluated the deployed file-system and compared it to a global storage system on a small-scale platform. Our experiments showed good performance and scalability for our method. We also proved the portability of our container-based method on another system equipped with a different storage technology.

We now plan to extend this work via three different ways. First, we would like to make the mechanism better integrated within job schedulers so allocation and deployment can be done without any user intervention. Second, we will come up with a finely configurable system for the deployment of data managers on different types of intermediate storage. Finally, we will continue to investigate the container-based approach and will propose a unique container packaging various data management systems required by a panel of applications (parallel file system, object-based storage, database, key-value store).

\section*{Acknowledgment}

This research is supported by the European project Maestro\footnote{\url{https://www.maestro-data.eu/}}.
The Maestro project has received funding from the European Union’s Horizon 2020 research and innovation program through grant agreement  801101.

\end{document}